\documentclass[9pt,twocolumn,twoside]{osajnl}
\usepackage{multirow}
\usepackage{amsmath}
\usepackage{graphicx}
\usepackage{hhline}
\usepackage{natbib}
\usepackage{colortbl}
\journal{ol} % Choose journal (ao, ol, josaa, josab)

\setboolean{shortarticle}{true} % true = letter, false = research article

\title{Dielectric Properties of Conductively Loaded Polyimides in the Far Infrared}
\author[1,*]{Kyle  R. Helson}
\author[1]{Kevin H. Miller}
\author[1]{Karwan Rostem}
\author[1]{Manuel Quijada}
\author[1]{Edward J. Wollack}

\affil[1]{NASA Goddard Space Flight Center, 8800 Greenbelt Road, Greenbelt, MD 20771, USA}
\affil[*]{Corresponding author: Kyle.Helson@nasa.gov}

\dates{Compiled \today}

\ociscodes{(160.4760) Optical properties; (310.3840) Materials and process characterization; (310.6188) Spectral properties.}

\doi{\url{http://dx.doi.org/10.1364/ol.XX.XXXXXX}}

\begin{abstract}
The dielectric properties of selected conductively-loaded polyimide samples are characterized in microwave through far infrared wavebands. These materials, belonging to the Vespel\textsuperscript{\textregistered} family, are more readily formed by direct machining than their ceramic loaded epoxy counterparts and present an interesting solution for realizing absorptive optical control structures. Measurements spanning a spectral range from 1 to 600\,${\rm cm^{-1}}$ (0.03 to 18\,THz) were preformed and used in parametrization of the media's dielectric function at frequencies below $\approx3\,$THz.
\end{abstract}

\setboolean{displaycopyright}{true}

\begin{document}
\maketitle
\thispagestyle{fancy}
\ifthenelse{\boolean{shortarticle}}{}{}

Conductively-loaded dielectric materials such as thermoplastics, castable epoxies, and paints have widespread use as absorbers in far-infrared applications~\cite{Smith1984,Persky1999}. In the microwave through far infrared, these materials can have a relatively large real component of the dielectric function and the geometry plays an important role in the construction of low reflectance radiometric flux calibration sources, waveguide terminations~\cite{Alison1987,Wollack2007,Jacob2018}, and baffling for stray light control~\cite{Sharp2012}. Manually mixing a variety of metals and semi-metals into epoxy formulations have achieved desirable absorptance performance. Typically, frozen premixed formulations are preferable due to their demonstrated homogeneity, consistency, and reduction of dielectric variability. Commonly available epoxies often contain fillers such as alumina, silica, or silicon-carbide particles to control the Coefficient of Thermal Expansion (CTE) and mitigate stresses between the loaded mixture and the substrate material upon thermal cycling. These CTE-matching additives are typically abrasive and increase wear experienced by tooling if subsequent machining of the material is required. 

Here the suitability of the Vespel\textsuperscript{\textregistered} SP polymer family~\cite{SP1IsoStatic} in these optical applications is considered. These materials are relatively easy to machine, can be formed with standard tooling for metals and plastics~\cite{VespelMachiningGuide}, and are widely employed for their mechanical and thermal properties. Polyimides are known to survive mechanical cycling to cryogenic temperatures and have potential optical applications if their dielectric properties are appropriately controllable and known. In this paper the dielectric functions of readily available polyimide mixture formulations are characterized in the microwave through far infrared. The polyimide mixture compositions investigated are provided in Table~\ref{tab:1}.

\begin{table*}
\centering
\resizebox{\linewidth}{!}{%
\begin{tabular}{|c|r *{3}{| c} |c| *{2}{|c} |l|}
\hline
  Sample &  Filling Fraction & Polyimide & Teflon\textsuperscript{\textregistered} & Graphite & MoS$_2$ & $\epsilon_{r}^\prime$ & $\epsilon_{r}^{\prime\prime}$  & Sample Description \\
 Media & {} & $[-]$ & $[-]$ & $[-]$ & $[-]$ & $[-]$ & $[-]$ & Fixture; Length \\ 
\hline 
\multirow{2}{*}{SP-1} & by Mass, $f_m$ &  1.000 & \cellcolor{gray!10} & \cellcolor{gray!10} & \cellcolor{gray!10} & 3.30 & 0.055 & Shim; 2.41\,mm\\
{} & by Volume, $f_v$ & 1.000 & \cellcolor{gray!10} &\cellcolor{gray!10} & \cellcolor{gray!10} & 3.28 & 0.051 & Split-Block; 50.8\,mm  \\ 
\hline
\multirow{2}{*}{SP-3} & by Mass, $f_m$ & 0.850 & \cellcolor{gray!10}   & \cellcolor{gray!10} & 0.150 & 3.47 &  0.098 & Shim; 2.42~mm\\
  & by Volume, $f_v$ & 0.952 & \cellcolor{gray!10} & \cellcolor{gray!10} & 0.048 & 3.42 &  0.086 & Split-Block; 32.8~mm \\ 
\hline
\multirow{2}{*}{SP-21} & by Mass, $f_m$ &  0.850 & \cellcolor{gray!10}  & 0.150 & \cellcolor{gray!10} & 10.8 & 0.76 & Shim; 2.42~mm \\
{}& by Volume, $f_v$ & 0.900 & \cellcolor{gray!10} & 0.100 & \cellcolor{gray!10} & 10.8 & 0.76 & Split-Block; 33.0~mm \\ 
\hline
\multirow{2}{*}{SP-211} & by Mass, $f_m$ &  0.750 & 0.100 & 0.150 & \cellcolor{gray!10} & 10.3 & 0.68 & Shim; 2.42~mm\\
{} & by Volume, $f_v$ & 0.824 & 0.071 & 0.104 & \cellcolor{gray!10} & 10.3 & 0.65 & Split-Block; 33.0~mm\\ 
\hline
\multirow{2}{*}{SP-22} & by Mass, $f_m$ &  0.600 & \cellcolor{gray!10} & 0.400 & \cellcolor{gray!10} & 22.5 & 2 & Shim; 2.42~mm \\
{} & by Volume, $f_v$ & 0.704 & \cellcolor{gray!10} & 0.296 & \cellcolor{gray!10} & 26 & 2.3 & Split-Block; 33.0~mm \\
\hhline{|=|=|=|=|=|=|=|=|=|}
\cellcolor{gray!10} & Density, $\rho$ $[\rm g/cm^3]$ & {1.43}~\cite{SP1IsoStatic} & {2.16~\cite{PTFEPropertiesHandbook}} & {2.27}~\cite{CRCHandbook} & {5.06}~\cite{CRCHandbook}& \cellcolor{gray!10} & \cellcolor{gray!10} &  \cellcolor{gray!10} \\
  \hline 
 
\end{tabular}
}
  \caption{Sample Composition and Measured Dielectric Properties at 1\,{$\rm cm^{-1}$} (30\,GHz). The mass filling fractions are based upon the nominal composition~\cite{VespelDesignHandbook} and the volume filling fractions are calculated from the bulk densities of the constituent materials~\cite{SP1IsoStatic,PTFEPropertiesHandbook,CRCHandbook}. Two measurements of the effective permittivities appear on the right hand side of the table for each sample type. The first entry indicates the dielectric parameters derived from a waveguide transmission line model~\cite{Wollack2008} using the shim data and the second corresponds to the E-plane split-block data. The parameters for each material agrees within experimental errors.}
  \label{tab:1}
\end{table*}

From an electromagnetic perspective these materials are dielectric mixtures realized from a polyimide host loaded with filler particles having differing conductivity. Dielectric mixing theory can provide useful insight into the optical properties of such composite absorber materials~\cite{Niklasson1981,Aspnes1982a,Aspnes1982b,Sihvola2008}. In the microwave, the size of the conductive particles in the medium is small compared to the wavelengths of interest. The fields are homogeneous and a mean field theory treatment is generally applicable in parameterizing the effective dielectric function of the material. In the far infrared, the wavelength of light is comparable to or smaller than the particle size and the response is commonly dominated by scattering processes. The simplifications enabled by considering these limits will be adopted in the analysis and presentation of the experimental observations that follow.

\begin{figure}[ht!]
	\centering
	\includegraphics[width=0.45\textwidth]{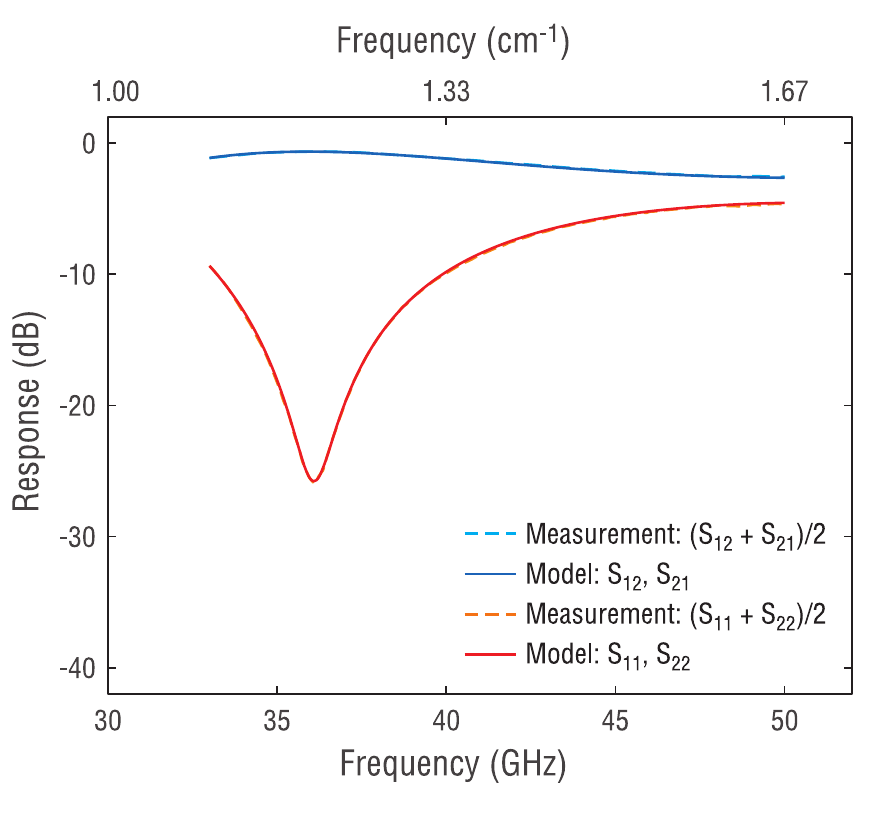}
	\caption{Measured VNA Reflection and Transmission Responses. The WR22.4 waveguide shim filled with SP-3 is provided as a representative data example. The average of the complex scattering amplitudes, $\rm S_{12}$ and $\rm S_{21}$, are used in estimating the measured power transmission (blue dashed lines). The average of $\rm S_{11}$ and $\rm S_{22}$ are used in estimating the measured power reflection (red dashed lines). Similarly, the scattering parameters computed from the transmission line model (solid lines) are over plotted in the figure to facilitate comparison. }
	\label{fig:VNA_data}
\end{figure}

For microwave characterization of the dielectric function a waveguide based metrology technique is employed. An E-plane waveguide split-block (length 33 or 50\,mm) and a shim (thickness 2.4\,mm) were used as fixtures to form Fabry-Perot waveguide resonators~\cite{Wollack2008}. The bulk material was directly machined by milling, press fit ({\it i.e.}, $\leq 20\, \mu\rm m$ clearance) into a WR22.4 rectangular waveguide, and the ends of each sample were lapped flush to the guide mounting flange surfaces. Two sample lengths were chosen to experimentally constrain the imaginary and real components of the dielectric function from the observed propagation in the guiding structures. The complex dielectric properties of the samples from 33 to 50\,GHz are extracted from the two-port scattering parameters measured with an Agilent PNA-X Vector Network Analyzer (VNA) using a Thru-Reflect-Line (TRL) calibration. Representative data acquired with the VNA and the modeled spectra appear in Figure~\ref{fig:VNA_data}. The extracted dielectric parameters from the waveguide measurements are provided in Table~\ref{tab:1}. The differences between the shim and split-block measurements of the same formulation arise from imperfections in preparation and sample fit within the differing waveguide fixtures.

The observed dielectric permittivity from the samples are then fit to a power-law logarithmic-mixing formula~\cite{Sihvola2008},
\begin{equation}
\hat{\epsilon}_{eff}^{\eta} = (1-f_{v}) \cdot \hat{\epsilon}_{host}^{\eta} + f_{v} \cdot \hat{\epsilon}_{fill}^{\eta} 
\label{eq:log_mixing}
\end{equation}
where the index is, $0 \leq \eta \leq 1$, and $f_{v}$ is the volume filling fraction of the medium loading the dielectric host. The value of $\eta$ is dependent on the detailed electromagnetic interaction between the particles in the mixture. In the limit the inclusions are isotropically distributed, the deviation in the constituents' permittivity is small compared to the mean permittivity, and the effective dielectric function is symmetric between the host and inclusion dielectric parameters, $\eta=1/3$ independent of the detailed internal structure~\cite{LandauLifshitz8}. In this form, Equation~\ref{eq:log_mixing} represents a  homogeneous dielectric mixture with separated-grains and is commonly referred to in the literature as the Looyenga formula~\cite{Looyenga1965}. 

\begin{figure}[ht!]
	\centering
	\includegraphics[width=0.45\textwidth]{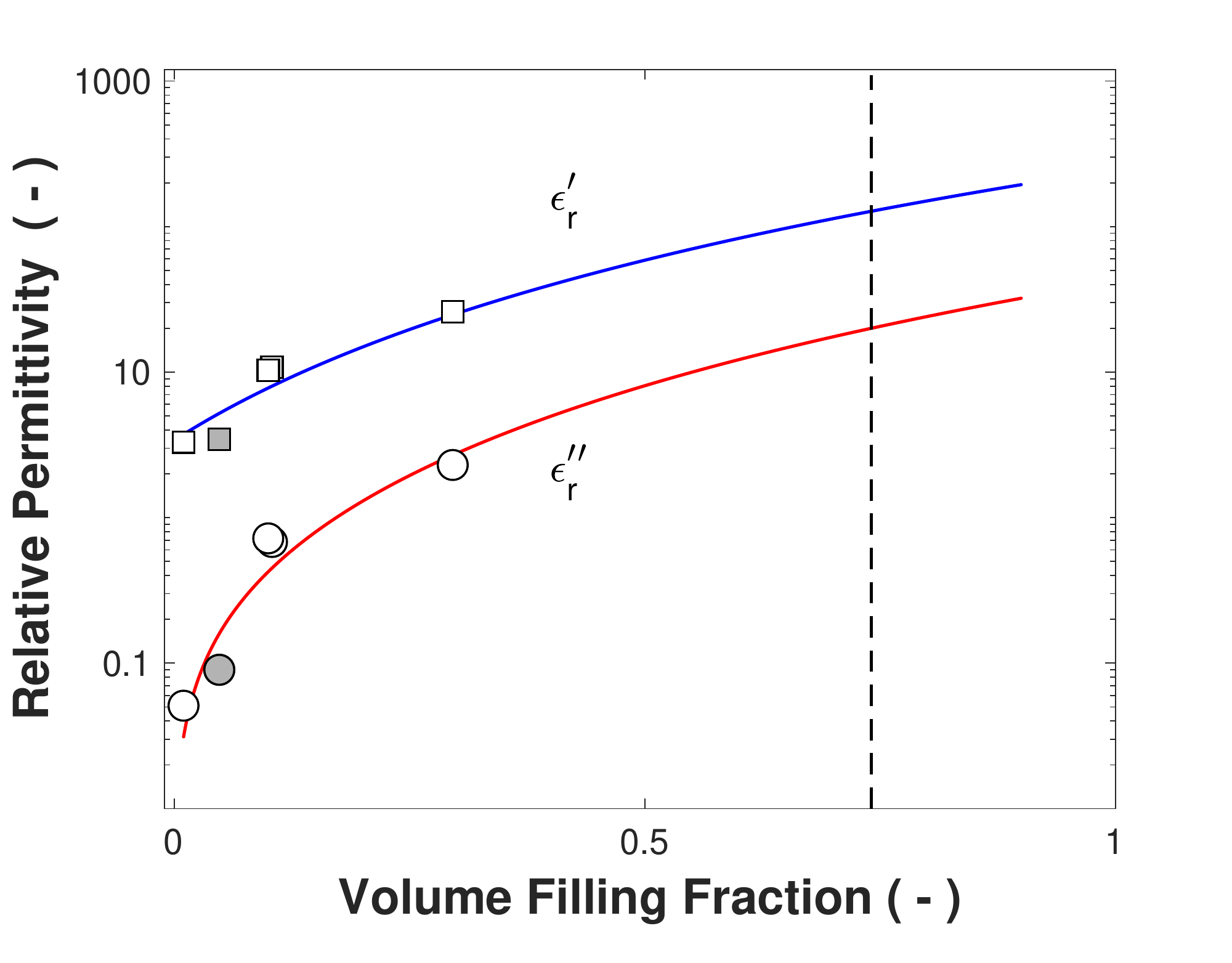}
	\caption{Relative Permittivity as a Function of Volume Filling Fraction. The squares and circles are respectively the real and imaginary components of the dielectric function. The symbol size reflects the magnitude of the systematic error in the derived parameters. The resulting complex dielectric parameters for the polyimide mixtures are fit to the logarithmic mixing formula as a function of $f_v$ of the conductively loading medium with $\eta \equiv 1/3$ (blue and red lines). The vertical dashed line indicates the maximum theoretically achievable volume filling fraction of a homogeneous mixture, which is a function of both the packing geometry and particle size distribution of the loading medium~\cite{Donev2004}. The MoS$_2$ loaded polyimide sample, SP-3, is indicated by grey symbols.}
	\label{fig:LossT}
\end{figure}

The observed effective dielectric function, $\hat{\epsilon}_{eff} \equiv \epsilon_r{^\prime} + i \cdot \epsilon_r^{\prime\prime}$, is derived from a joint least-squares fit to the waveguide shim and split-block VNA data for each sample type~\cite{Janezic1999}. See Fig.~\ref{fig:LossT}. It is important to note that $\hat{\epsilon}_{eff}$ has a degeneracy as a function of $f_{v}$ and dielectric parameters, which is not highly constrained over the limited parameter range of commercially available samples. The dielectric function of the conductive loading medium is estimated as $\hat{\epsilon}_{fill} = 246+42i$, assuming $\hat{\epsilon}_{host} = 3.4 + 0.051i$ as the magnitude of the polyimide host. The inferred bulk dielectric properties of graphite~\cite{DraineLee1984} and polyimides ({\it e.g.}, Kapton\textsuperscript{\textregistered}~\cite{Smith1975}, Cirlex\textsuperscript{\textregistered}~\cite{Lau2006}, {\it etc.}) derived from these mixtures are consistent with prior observations at millimeter and sub-millimeter wavelengths. At the relatively low concentration employed here, MoS$_2$ has a similar influence on the properties of the end dielectric mixture as graphite, as can be seen in Fig.~\ref{fig:LossT}. Dilution of the polyimide host with Teflon\textsuperscript{\textregistered} ($\hat{\epsilon} = 2.1 + 0.0002i$) slightly lowers the effective dielectric function of SP-211 relative to SP-21 consistent with theoretical expectations.

\begin{figure}[ht!]
	\centering
	\includegraphics[width=0.45\textwidth]{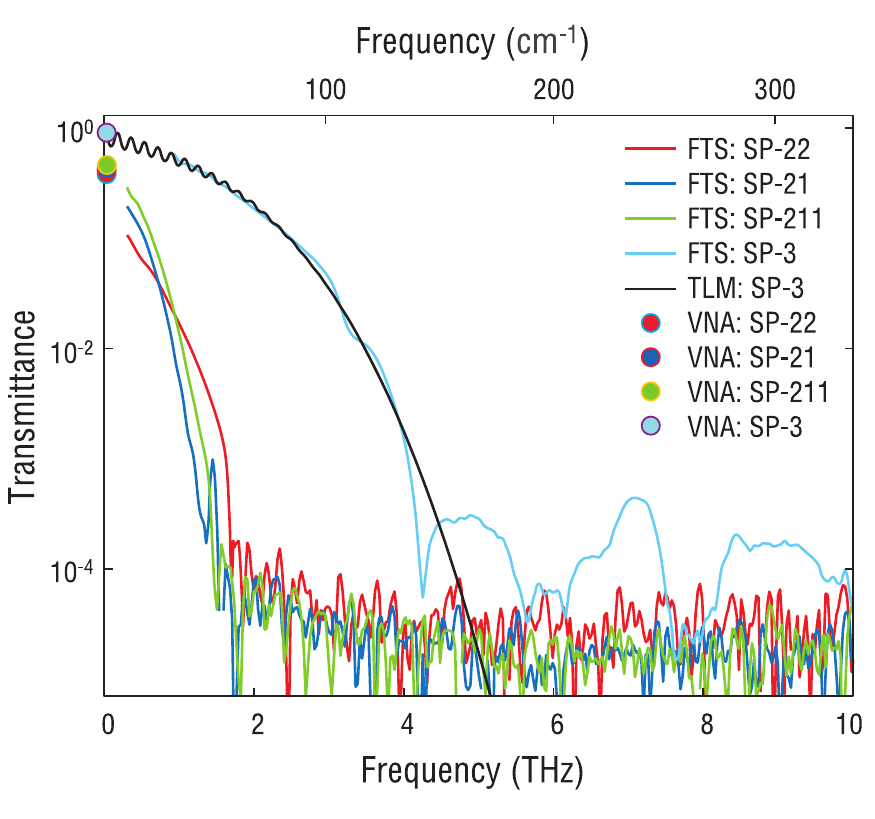}
	\caption{Measured FTS Transmittance Summary. Circular symbols are calculated from the FTS sample thickness and complex permittivity, ${\hat \epsilon}_{eff}$, extracted from the VNA spectra. Colored curves show FTS data for each of the different samples. Knowledge of transmittance above $1.8\,$THz for samples SP-21, SP-211, and SP-22 is limited by the noise floor, $\approx 1 \times 10^{-4}$, set by the observation signal-to-noise and averaging of FTS data scans. The black curve is the TLM fit for the SP-3 sample transmission response. Leakage through this sample was observed at $>4.4\,$THz due to the mixture's lower dielectric function and absorption.}
	\label{fig:FTS_data}
\end{figure}

The materials were also characterized with a Bruker IFS 125 high-resolution Fourier Transform Spectrometer (FTS) from 10 to 600\,{$\rm cm^{-1}$} (0.3 to 18 THz). The optical samples were machined and lapped to a plane-parallel geometry to enable illumination by the FTS beam by an area up to 10\,mm in diameter. The thicknesses were measured with a digital micrometer and are indicated in~Table~\ref{tab:2}. The FTS was used in a focused (f/6.5) beam configuration for characterization of the samples with a Si-doped 4\,K LHe bolometer. In practice, the illuminated sample area is $\sim$~6\,mm in diameter. To cover the desired spectral range two beam splitters were used ({\it i.e.}, 50-\textmu m mylar beamsplitter covering 10--100~cm$^{-1}$ and 6-\textmu m Si-coated mylar beamsplitter from 30--600~cm$^{-1}$) and the transmittance from each band agreed to within 0.5\% in their region of overlap. The datasets were merged into a single spectrum by computing a weighted average (Fig.~\ref{fig:FTS_data}).

The observed optical response is modeled as a non-magnetic medium with a constant effective homogeneous dielectric function, $\hat{\epsilon}_{eff}$. The measured sample thicknesses are adopted as the physical section lengths in a air-dielectric-air layered Transmission Line Model (TLM)~\cite{Yeh1988}. The propagation  constant, $\gamma = (\alpha + i \cdot \beta) + \alpha_{s}$, has contributions from both intrinsic and extrinsic losses in the effective medium. The material's attenuation and phase constants:
\begin{equation}
\alpha = \kappa \cdot \frac{2\pi}{\lambda_o} =
\left[{ \frac{1}{2} \left( \sqrt{\epsilon_r'^2+\epsilon_r''^2} 
- \epsilon_r' \right) } \right]^{1/2}  
\cdot \frac{2\pi}{\lambda_o} 
\end{equation}
\begin{equation}
\beta = n \cdot \frac{2\pi}{\lambda_o} =
\left[{ \frac{1}{2} \left( \sqrt{\epsilon_r'^2+\epsilon_r''^2} 
+ \epsilon_r' \right) } \right]^{1/2}  
\cdot \frac{2\pi}{\lambda_o} 
\end{equation}
are evaluated with the measured relative dielectric function, $\hat{\epsilon}_{eff}(1\,{\rm cm^{-1}})$. Here $\lambda_o$ and $\hat{n}=n+i\kappa$ are respectively the free-space wavelength and the refractive index of the medium. Following Halpern {\it et al.} \cite{Halpern1986}, the extrinsic scattering loss in the medium is modeled as a power-law,
\begin{equation}
2 \alpha_{s}(\nu) = a_{o} \cdot (\nu/[1\,{\rm cm^{-1}}])^{b},
\end{equation}
where $\nu$ is the frequency in inverse centimeters and the parameters are derived from a least-squares fit to the FTS data. See Table \ref{tab:1}. A representative example of the modeled response is included in the transmittance plotted in Figure~\ref{fig:FTS_data}.

\begin{table}[ht!]
\centering
\begin{tabular}{|l|c|c|c|}
\hline
Sample & $a_{o}$ & $b$ & Thickness \\
Media & [Np/$\mu$m] & $[-]$ & $[\mu\rm m]$ \\
\hline
SP-3   & $4.1\times 10^{-9}$ & 3.0 & 452 \\
\hline
SP-21  & $3.4\times 10^{-6}$ & 2.1 & 452 \\
\hline
SP-211 & $3.0\times 10^{-6}$ & 2.1 & 426 \\
\hline
SP-22  & $1.5\times 10^{-5}$ & 2.0 & 102 \\
\hline
\end{tabular}
\caption{Polyimide Mixture Attenuation Parameter Summary. The thicknesses indicated correspond to the measured values for the data presented in Figure~\ref{fig:FTS_data}. The number of significant figures reflects the observed parameter uncertainty in repeated FTS measurement trails and between samples with the same composition but having differing total thickness.}
\label{tab:2}
\end{table}

An interesting application for this class of materials is the construction of low-reflectance waveguide terminations and compact calibration targets for the far infrared. To explore this possibility a bulk rod of SP-22 was machined into a pair of approximately conical circular terminations with full taper angles of $\theta \approx 7^{\circ}$ and $ 10^{\circ}$ following the general design approach outlined in~\cite{Wollack2007}. An absorber insert with outer diameter of $2.9$\,mm allows insertion and use at 100 GHz in circular waveguide housing (internal diameter $2.97$\,mm, length $25.4$\,mm) matched to the cutoff frequency of WR10.0~\cite{Pyle1964}. The waveguide fixture accommodates the sample's entire ($\le 15$\,mm) length during VNA characterization. Waveguide simulators with similar construction and appropriate symmetry can be utilized for characterizing the angular response of infinite array tilings~\cite{Hansen1998} and are of value in prototyping free-space calibration targets~\cite{Fixsen2006,Chuss2017}.

\begin{figure}[ht!]
	\centering
	\includegraphics[width=0.45\textwidth]{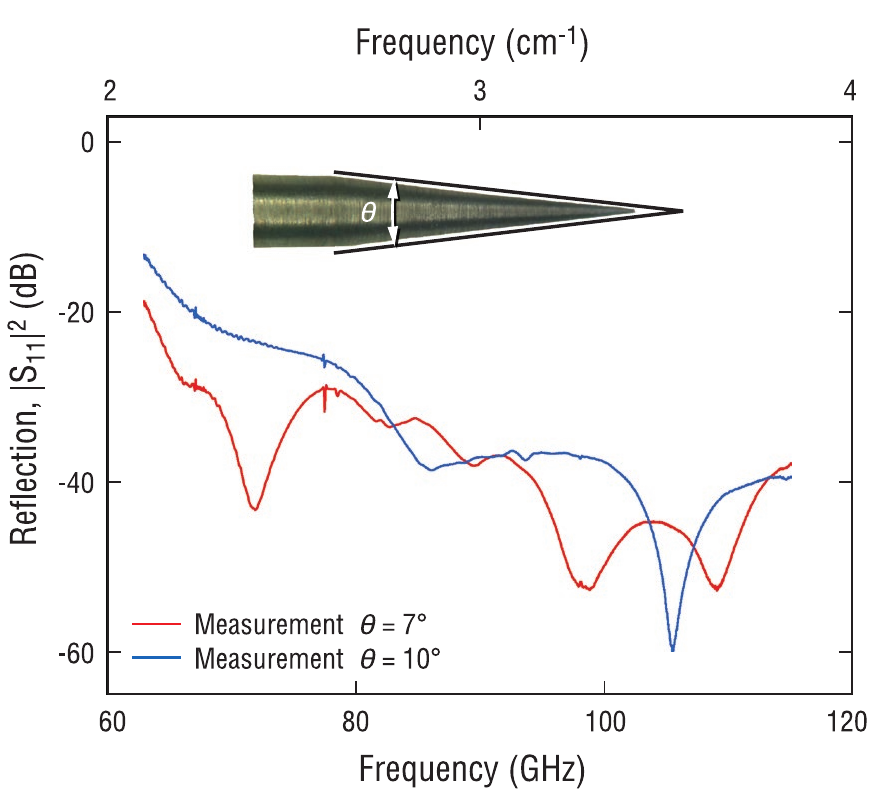}
	\caption{Measured SP-22 Waveguide Termination Reflection Response. The $\theta \approx 10^{\circ}$ and $7^{\circ}$ taper angle data are designated by blue and red solid lines respectively. The inset depicts an example of the tapered load geometry with $\theta \approx 7^{\circ}$. The cone full angle is indicated by solid back lines. The sample is inserted into a circular waveguide housing and the reflectance is measured with a VNA.}
	\label{fig:WG_load}
\end{figure}

The measured absorber insert tip diameter, $\approx30\,\mu{\rm m}$, is smaller than readily achievable through molding or grinding a typical loaded epoxy medium such as MF117 and related formulations~\cite{Rostem2013}. The achieved profile deviates slightly from the targeted conical form due to the stiffness of the material and its uncompensated response during mechanical forming on a lathe. If an alternative to direct machining is necessitated by the component shape, size, or quantity, the medium could potentially be formed by direct forming or molding. Reflections from the waveguide loads were measured using a one-port vector reflectometry technique~\cite{Rostem2013}. Over the measured spectral range the envelope of the peak reflectance for the loads are qualitatively similar. See Figure~\ref{fig:WG_load}. Reducing the taper angle largely mitigates reflections near guide cutoff. These loaded materials could also find application as attenuators, edge-filters, and thermal-blockers for the control of broadband radiation. The reported dielectric parameters for conductively loaded polyimides can be applied in applications where a knowledge of the far infrared optical response of these materials is required.

%\section*{Acknowledgments}
\noindent {\large\textbf{Acknowledgment.}} The authors thank B.T. Bulcha, S. Pedek, and P. Cursey for their respective contributions to material characterization and sample preparation.

%\section*{Funding}
\noindent {\large\textbf{Funding.}} 
This work was funded under a NASA Research Opportunities in Earth and Space Science (ROSES) Strategic Astrophysics Technology (SAT) award and the NASA Postdoctoral Program (NPP).\\

%\pagebreak
% Informational fifth page
\vspace{-1mm}

\bibliography{Bibliography}
\bibliographyfullrefs{Bibliography}
\end{document}